\documentclass[conference]{IEEEtran}
\IEEEoverridecommandlockouts
\usepackage{cite}
\usepackage{amsmath,amssymb,amsfonts}
\usepackage{algorithmic}
\usepackage{graphicx}
\usepackage{multirow,booktabs}
\usepackage{subfigure,stfloats}
\usepackage{textcomp}
\usepackage{xcolor}
\usepackage{hyperref}
\usepackage{epstopdf}

\def\BibTeX{{\rm B\kern-.05em{\sc i\kern-.025em b}\kern-.08em
    T\kern-.1667em\lower.7ex\hbox{E}\kern-.125emX}}
\begin{document}

\newcommand{\theHalgorithm}{\arabic{algorithm}}
\newcommand{\shigang}[1]{\textcolor{red}{[Shigang: #1]}}


\title{DeFuzz: Deep Learning Guided Directed Fuzzing}

\author{\IEEEauthorblockN{1\textsuperscript{st} Xiaogang Zhu}
\IEEEauthorblockA{
\textit{Swinburne University Of Technology}\\
Melbourne, VIC, Australia \\
xiaogangzhu@swin.edu.au}
\and
\IEEEauthorblockN{2\textsuperscript{nd} Shigang Liu}
\IEEEauthorblockA{
\textit{Swinburne University Of Technology}\\
Melbourne, VIC, Australia \\
shigangliu@swin.edu.au}
\and
\IEEEauthorblockN{3\textsuperscript{rd} Xian Li}
\IEEEauthorblockA{
\textit{Swinburne University Of Technology}\\
Melbourne, VIC, Australia \\
xianli1990@outlook.com}
\and
\IEEEauthorblockN{4\textsuperscript{th} Sheng Wen}
\IEEEauthorblockA{
\textit{Swinburne University Of Technology}\\
Melbourne, VIC, Australia \\
swen@swin.edu.au}
\and
\IEEEauthorblockN{5\textsuperscript{th} Jun Zhang}
\IEEEauthorblockA{
\textit{Swinburne University Of Technology}\\
Melbourne, VIC, Australia \\
junzhang@swin.edu.au}
\and
\IEEEauthorblockN{6\textsuperscript{th} Camtepe Seyit}
\IEEEauthorblockA{
\textit{CSIRO \& DATA61}\\
Marsfield, NSW, Australia \\
seyit.camtepe@data61.csiro.au}
\and
\IEEEauthorblockN{6\textsuperscript{th} Yang Xiang}
\IEEEauthorblockA{\textit{Swinburne University Of Technology}\\
Melbourne, VIC, Australia \\
yxiang@swin.edu.au}
}

\maketitle

\begin{abstract}
Fuzzing is one of the most effective technique to identify potential software vulnerabilities. Most of the fuzzers aim to improve the code coverage, and there is lack of directedness (e.g., fuzz the specified path in a software). In this paper, we proposed a deep learning (DL) guided directed fuzzing for software vulnerability detection, named DeFuzz. DeFuzz includes two main schemes: (1) we employ a pre-trained DL prediction model to identify the potentially vulnerable functions and the locations (\textit{i.e.}, vulnerable addresses). Precisely, we employ Bidirectional-LSTM (BiLSTM) to identify attention words, and the vulnerabilities are associated with these attention words in functions. (2) then we employ directly fuzzing to fuzz the potential vulnerabilities by generating inputs that tend to arrive the predicted locations. To evaluate the effectiveness and practical of the proposed DeFuzz technique, we have conducted experiments on real-world data sets. Experimental results show that our DeFuzz can discover coverage more and faster than AFL. Moreover, DeFuzz exposes 43 more bugs than AFL on real-world applications.
\end{abstract}


\section{Introduction}
Computer software is a crucial part of the modern world. The software makes the world become smart than before. However, cyber threats attributed to vulnerabilities in software are becoming a serious security problem \cite{liu2019deepbalance}. For example, it is estimated that it will cost \$6 trillion to combat cybercrime by 2021, which is double of the cost in 2015. Therefore, the vulnerabilities in the software should be ideally identified and fixed before the software get deployed. 

Many algorithms have been developed for software vulnerability detection \cite{coulter2020code}, \cite{ghaffarian2017software}, \cite{lin2020software}. Among these techniques, fuzzing is one of the most effective approach to identify potential software vulnerabilities \cite{zhu2019feature}. However, fuzzing has the problem of time-consuming because it identify bugs using randomly generated inputs. Fuzz testing usually hard to arrive all the code coverage given a real-world scenario. For example, it is very hard to generate inputs for some sanity checks with the increasing size of the real-world programs \cite{peng2018t}. Moreover, previous study showed that vulnerable functions usually come from a small part of the whole program. To address these problems, directed greybox fuzzing, which uses the inputs that are generated with the objective of reaching specific locations, has been developed. Nevertheless, greybox fuzzing usually has the problem that it cannot be effectively directed \cite{bohme2017directed}. Therefore, there is still necessary to research into the directed fuzzing, which can fuzz the specific targeted locations with less of resources.

In this work, we take the advantages of deep learning \cite{ban2019performance}, and developed a deep learning directed fuzzing for bugs identification. Deep learning has shown successful in software vulnerability detection \cite{li2018vuldeepecker}, \cite{liu2019cyber}, specifically in learning the high-level representations for both vulnerable and not-vulnerable programming features. The high-level representations usually contains richer useful information than the generic hand-crafted features \cite{lin2018cross}. However, deep learning-based techniques usually has high false positives (e.g., more than 20\%) \cite{lin2018cross}, \cite{li2018vuldeepecker}, \cite{liu2019cyber}. In this work, we employ deep learning to identify the potential vulnerable functions and the vulnerability locations. Then we run the directed fuzzing to fuzz the vulnerable locations to reduce the false positive rate. 

Specifically, given a software program, we first extract functions from the program, and then we extract Abstract Syntax Trees (ASTs) for each function. The ASTs we collected will be used as the training samples. The training samples are from LibTIFF, LibPNG, and FFmpeg, which are three open-source projects.
We label the vulnerable functions by following the Common
Vulnerability and Exposures and the National Vulnerability Database. The assumption is that if there are at least one vulnerability in the function, then the function will be labeled as vulnerable, otherwise, it will be labeled as not vulnerable. Then, we train an attention-based deep learning model using the ground truth. The pre-trained model will be applied to the test cases (i.e., a piece of program or a software program). The outputs will be the locations of the potential vulnerabilities (the vulnerabilities are associated with these attention words in functions). 
Afterwards, we fuzz the vulnerable locations using directed fuzzing \cite{Boehme2017Directed}. We call this as deep learning guided directed fuzzing, namely DeFuzz. 
The main contributions of this work are as follows:
\begin{itemize}
\item We propose to employ attention-based deep learning to highlight code keyword, which will be used to identify the penitential vulnerable locations (i.e., vulnerable addresses) for real-world security problems.  

\item We propose to use deep learning to guide the directed fuzzing. When deep learning can successfully predict the exact or the neighbours of vulnerability locations, directed fuzzing is efficient to detect vulnerabilities.

\item We conducted experiments on real-world scenarios. Experimental results show that the proposed DeFuzz can discover more coverages and faster than AFL. For example, DeFuzz exposes 43 more bugs than AFL on real-world applications.

\end{itemize}

\section{Related work}
In this section, we briefly review of the related work from two perspectives: machine learning-based vulnerability detection and fuzz testing for software vulnerability discovery. For more information, please refer to \cite{coulter2020code}, \cite{ghaffarian2017software}, \cite{lin2020software}, \cite{li2018fuzzing}, \cite{manes2019art}. 

\subsection{Machine Learning based vulnerability detection}
Machine learning has been widely used for software vulnerability detection. Morrison et al. [36] investigate the machine learning techniques for software vulnerability identification, their study show that machine learning techniques is useful in vulnerability detection. However, the classification performance based on source code usually performs better than binary code level.

Machine learning techniques for software vulnerability detection can be divided into two stages: feature learning, and model building. Feature learning is a case by case process. For example, 
Shar and Tan \cite{shar2013predicting} propose to consider static code attributes for model building to predict specific program statements for SQL injection and cross site scripting.
Yamaguchi et al. \cite{yamaguchi2015automatic} propose to use clustering algorithms on code property graph based on C source code for taint-style vulnerability discovery. Alves et al. \cite{alves2016experimenting} have done an experimental study over machine learning techniques, their experimental results show that random forest usually has a better performance. 
DisovRE \cite{eschweiler2016discovre} uses control flow graph for function similarity match to identify potential bugs given a piece of software. VDiscover \cite{grieco2016toward} employs lightweight static and dynamic features to identify potential vulnerabilities given a test case. SemHunt \cite{li2017semhunt} proposes a scheme to predict vulnerable functions and their pairs of patched and unpatched functions using the binary executable programs.
Calzavara et al. \cite{calzavara2019mitch} propose to consider HTTP requests as features and employ supervised machine learning to detect Cross-Site Request Forgery vulnerabilities.

Meanwhile, deep learning has been widely accepted and applied for software vulnerability detection. This is because deep learning has the capability of learning high-level feature representations given a program.
Lin et al. \cite{lin2019software} propose to employ LSTM for function-level vulnerability detection based on ASTs. Li et al. \cite{li2018vuldeepecker} consider BiLSTM using code gadgets as features for software vulnerability detection.  
Liu et al. \cite{liu2019deepbalance} realize that there is class imbalance problem \cite{liu2017fuzzy} in software vulnerability detection, and propose a fuzzy-based oversampling algorithm using ASTs for vulnerability detection. Furthermore, a deep learning based cross domain software detection has been developed \cite{liu2020cd}, this study employs metric transfer learning framework to minimize the distribution differences between the source domain and target domain.
Liu et al. \cite{liu2019cyber} propose to employ attention-based deep learning using binary instruction as features for binary-level vulnerability detection. 
For more information about software vulnerability detection using machine, please refer to \cite{xue2019machine}.

\subsection{Directed Fuzzing}
Coverage-guided fuzzing is a widely-used fuzzing solution due to its effectiveness \cite{Aschermann2020IJON, Gan2020GREYONE, Nagy2018Full, Chen2019Matryoshka, Lyu2019MOPT, Aschermann2019REDQUEEN, csi2020zhu, She2019NEUZZ, Stephens2016Driller, Gan2018CollAFL}. Coverage-guided fuzzing aims to cover as much coverage as possible. The assumption is that with more coverage discovered, it has higher chance to expose bugs. Driller \cite{Stephens2016Driller} uses symbolic execution to solve path constrains, which is the major road rock to discover more coverage, for fuzzing. CollAFL \cite{Gan2018CollAFL} improves the coverage via resolving edge collision. 
Angora \cite{Chen2018Angora} and Matryoshka \cite{Chen2019Matryoshka} regard a single path constraint as a black-box function, and use gradient descent to bypass the path constraint. As some path constraints compare variables with constant values, REDQUEEN \cite{Aschermann2019REDQUEEN} uses the constant values to bypass constraints. The mutation of the entire input for a program requires much time. Therefore, GREYONE \cite{Gan2020GREYONE} first finds the relation between input bytes and path constraints, and then only mutates the related bytes. This solution largely reduce the searching space of input. As the random nature of fuzzing, the execution speed is critical for fuzzing. Full-speed fuzzing \cite{Nagy2018Full} improves the execution speed of fuzzing via only tracing execution that discovers new coverage. Although coverage-guided fuzzing has achieved a great success, it has problems to detect vulnerabilities efficiently. The reason is that, coverage-guided fuzzing spends almost the time for each path, wasting time for the non-vulnerability paths.

To improve the efficiency of coverage-guided fuzzing, directed fuzzing aims to cover the potential bug locations, which is more effective than coverage-guided fuzzing.
Directed fuzzing is one of the most promising fuzzing solutions to detect vulnerabilities. Directed fuzzing instruments some potential bug locations and guides fuzzing to reach the locations.
AFLGo \cite{Boehme2017Directed} utilises the commit information from Github to set bug locations. Then, it assigns more mutation time for the execution paths that are closer to these locations. SAVIOR \cite{Chen2020SAVIOR} sets the potential bug locations using Undefined Behavior Sanitizer (UBSan). UBSan can detect variables that are used before definition. 
Besides UBSan, other sanitizers can also be used to set the potential bug locations. ParmeSan \cite{Oesterlund2020ParmeSan} regards the sanitization instrumentation as the target locations and guides fuzzing to these insteresting basic blocks.
To discover memory corruption vulnerabilities, TortoiseFuzz \cite{Wang2020Not} statically analyses locations of sensitive memory operations, and then guides fuzzing to these memory related locations. Memory corruption is a common vulnerability among all kinds of vulnerabilities. However, in this paper, we utilise the results from deep learning to set the potential bug locations.

\begin{figure*}[!t]
\centering
\includegraphics[trim = 0mm 0mm 0mm 10mm, scale=0.72]{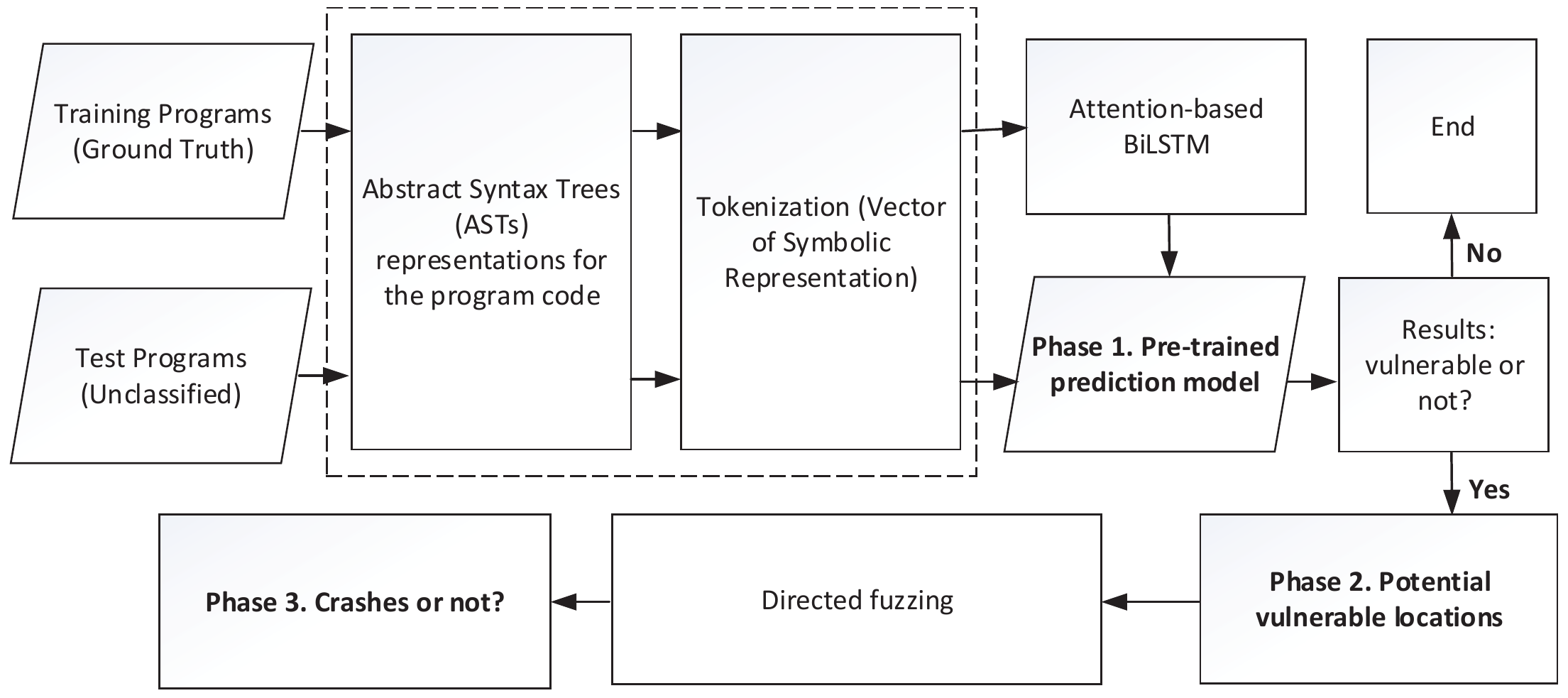}
\caption{Framework of the proposed DeFuzz. There are three phases: prediction model training, potential vulnerable location identification, and directed fuzzing. 
}
\label{Figframework}
\end{figure*}

\begin{figure}[!t]
\centering
\includegraphics[scale=0.5]{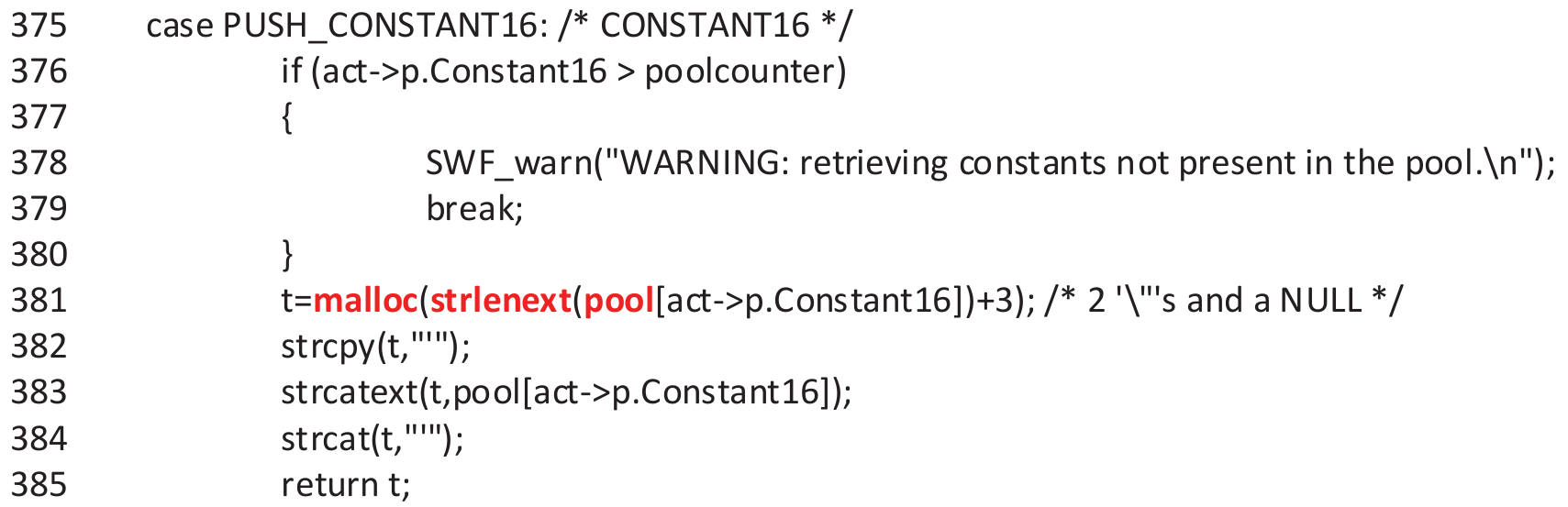}
\caption{Location of potential vulnerability identified from a piece of code of (or subroutine of) getString() function from decompile.c file.
}
\label{FigExample}
\end{figure}

\section{Proposed DeFuzz}
In this section, we discuss the main idea of DeFuzz. DeFuzz is a deep learning guided directed fuzzing for software vulnerability identification. Fig.\ref{Figframework} is the framework of the proposed DeFuzz.
As shown in Fig.\ref{Figframework}, one can see there are three phases in the proposed scheme: 1) Prediction model training; 2) potential vulnerable location identification; 3) fuzz the potential vulnerable locations using directed fuzzing.

In the first phases, we train the attention-based deep learning model based on the ground truth we collected. This model is the pre-trained prediction model as we can see from Fig.\ref{Figframework}.
In this study, the ground truth data samples are C/C++ functions \footnote{Function means a group of statements that together perform a task.} with labels which means the samples will be labeled as either vulnerable or not vulnerable \cite{liu2020cd}, \cite{lin2019software}. The data samples are from three open-source projects including FFmpeg, LibPNG, and LibTIFF \footnote{Download from: \url{https://github.com/cybercodeintelligence/CyberCI.}}. We collect the sources from Github, and then we manually labeled 417 vulnerable functions and 6,860 not-vulnerable functions.
The assumption is that if there is at least one vulnerability identified in the function, then the function will be labelled as vulnerable, otherwise, it will be labeled as not vulnerable. Then, we extract ASTs representations for the program code at function level (a detailed description of ASTs can be found in \cite{liu2019deepbalance}, \cite{lin2018cross}).  
The ASTs are tokenized into AST sequences before they are fed to the the attention-based Bidirectional LSTM (BiLSTM) for feature representation learning. 
Finally, the fully connected neural network will be employed to train the prediction model based on the feature vector representations.

\begin{table}[!tbp]
  \centering
  \caption{Crashing Trace of Listaction\_d. Fuzzing has to generate an input that exercises all the ten listed functions.}
    \begin{tabular}{ll}
    \toprule
    Functions in a Crashing Trace & File \& Line \\
    \midrule
    main  & main.c:354 \\
    readMovie & main.c:281 \\
    outputBlock & outputtxt.c:2933 \\
    outputSWF\_DOACTION & outputtxt.c:1620 \\
    decompile5Action & compile.c:3517 \\
    decompileActions & decompile.c:3494 \\
    decompileAction & decompile.c:3413 \\
    decompileSingleArgBuiltInFunctionCall & decompile.c:2994 \\
    newVar\_N & decompile.c:725 \\
    getString & decompile.c:381 \\
    \bottomrule
    \end{tabular}%
  \label{Cranshing_Trace}%
\end{table}%

The second phase is used to identify potential vulnerabilities given any test cases. The inputs are C or C++ programs. The outputs are the vulnerable locations at function level (i.e., the test cases are programs at function level, the results will be the locations of potential vulnerabilities regarding each function). It is worth to note that the location of the vulnerabilities is identified by using the attention words produced by the attention-based BiLSTM model. We consider attention-based deep learning to train the prediction model because the attention words produced by the model can be helpful for us to identify the location of the vulnerabilities.  Fig.\ref{FigExample} gives an example, we extracted $ getString() $ from $ decompile.c $ file of the \texttt{libming} project as a test case for the pre-trained model. When we fed this function to our pre-trained model for testing. The output of the attention words are: $ malloc $, $ Constant16 $, $ strlenext $ and $ pool $. After we manually check with $ getString() $ function. We then identified the potential vulnerability is from the line 381 of $ t=malloc(strlenext(pool[act-> p.decomstant16])+3) $.

In the third phase, we employ the directed fuzzing to fuzz the vulnerable locations identified in the second phase. Specifically, we use AFLGo as the directed fuzzing. When AFLGo gets the potential vulnerability locations, which are generated by deep learning, AFLGo calculates distance between the current execution path and the target locations (Details can be found in \cite{Boehme2017Directed}). Then, AFLGo assigns mutation chances based on the distance, \textit{i.e.}, the smaller the distance is, the more mutation chances the path is assigned. Therefore, directed fuzzing spends more time on execution paths that examine target locations, which improves the efficiency of detecting vulnerabilities. Note that, although we use AFLGo to demonstrate the efficacy of our DeFuzz, we will use other directed fuzzers or design our own one in the future.
As an example, Table \ref{Cranshing_Trace} shows the bug location in the 381st line of the file \texttt{decompile.c} in program \texttt{libming}, and the bug is exposed running application \texttt{listaction\_d}. As shown in Table \ref{bugs} (refer to Section \ref{Section:Experiments}), deep learning precisely predicts this bug location, and fuzzing is guided to this bug location. Starting from the function \textit{main()}, fuzzing has to generate an input that exercises another nine functions and trigger the bug in function \textit{getString()}. The 381st code line of function \textit{getString()} calls \textit{malloc()}, which causes a segmentation fault.


\section{Experiments}
\label{Section:Experiments}

We evaluate our DeFuzz on eight real-world open-sourced programs, which are shown in Table \ref{crashes_num}. 
The eight target programs include seven programs from \texttt{libming} \cite{libming} library, which deals with \texttt{.swf} files, and \texttt{gifsponge} from \texttt{gif}, which manipulates GIF images. The parameters used for the eight programs are shown in the fifth column of Table \ref{crashes_num}.
These applications are widely used in the real world, and their security will have a significant impact on the applications developed based on them. Therefore, we chose them as our target programs.
We compare our DeFuzz to AFL with the same configure, \textit{i.e.}, the same initial input and timeout. We set the timeout as 24 hours, \textit{i.e.}, the execution will terminate when fuzzing runs more than 24 hours.
We run the experiments on the computer with \textit{AMD Ryzen Threadripper 2990WX 32-core Processor 128GB RAM}. Each program runs on a single core.
We will evaluate the performance of our DeFuzz in terms of execution speed, code coverage, and bug discovery.

\begin{table}[!t]
  \centering
   \caption{The Number of Unique Crashes. DeFuzz detects more crashes than AFL.}
   
    \begin{tabular}{|l|c|c|c|c|}
    \hline
    Application  & Version & DeFuzz & AFL & Parameters  \\
    \hline
    listaction\_d & \multicolumn{1}{c|}{\multirow{7}{*}{commit 50098}} & 10    & 16 &  \multicolumn{1}{c|}{\multirow{8}{*}{@@}}\\ 
\cline{1-1}\cline{3-4}    listswf\_d &       & 162   & 216 &\\
\cline{1-1}\cline{3-4}    swftocxx &       & 13    & 0 &\\
\cline{1-1}\cline{3-4}    swftoperl &       & 87    & 0 &\\
\cline{1-1}\cline{3-4}    swftophp &       & 166   & 0 &\\
\cline{1-1}\cline{3-4}    swftopython &       & 155   & 0 &\\
\cline{1-1}\cline{3-4}    swftotcl &       & 165   & 0 &\\
\cline{1-4} gifsponge & commit 72e1f & 11    & 0 &\\
    \hline
    \end{tabular}%
  \label{crashes_num}%
\end{table}%

\subsection{Excution Speed}
	
\begin{figure}[!t]
	\centering
		\begin{minipage}[t]{0.9\columnwidth}
			\centering
	        \includegraphics[width=\columnwidth]{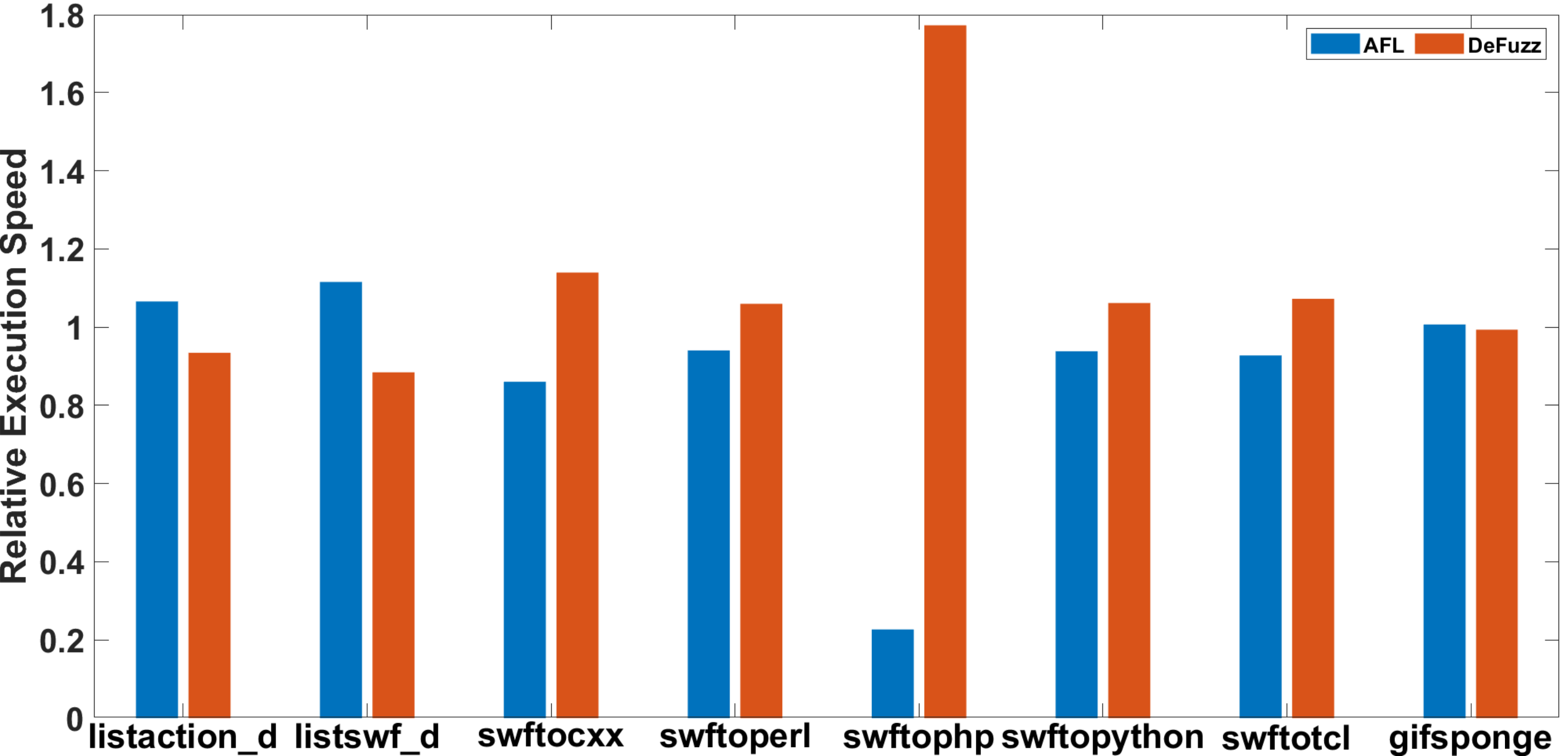}
		\end{minipage}%
	\caption{Relative Execution Speed. The relative execution speed is the execution speed of each fuzzer compared to the average execution speed. A higher relative execution speed indicates more test cases being examined in the same time. The execution speed of DeFuzz is close to AFL's.}
	\label{relative_speed}%
\end{figure}

\begin{table}[htbp]
  \centering
  \caption{The Number of Real-world Bugs And Types. DeFuzz Discovers More Bugs Than AFL.}
    \begin{tabular}{lll}
    \toprule
    Applications  & DeFuzz & AFL \\
    \midrule
    \multicolumn{3}{c}{Detected Bugs per Project} \\
    \midrule
    listaction\_d & 5     & 6 \\
    listswf\_d & 12    & 14 \\
    swftocxx & 6     & 0 \\
    swftoperl & 8     & 0 \\
    swftophp & 10    & 0 \\
    swftopython & 8     & 0 \\
    swftotcl & 12    & 0 \\
    giflib & 2     & 0 \\
    \midrule
    \multicolumn{3}{c}{Detected bugs Type} \\
    \midrule
    out-of-memory & $\surd$     & $\surd$ \\
    double free & $\surd$     & $\times$ \\
    \midrule
    Total & 63    & 20 \\
    \bottomrule
    \end{tabular}%
  \label{bugs-type}%
\end{table}%

\begin{table}[htbp]
  \centering
  \caption{The Time Of Detecting First Crashes By AFL. Over 24 Hours, AFL detects the First Crashes.}
    \begin{tabular}{|l|l|}
    \hline
    Application & Time (h) \\
    \hline
    swftocxx & 26.8  \\
    \hline
    swftoperl & 25.7  \\
    \hline
    swftophp & 27.4  \\
    \hline
    swftopython & 25.5  \\
    \hline
    swftotcl & 26.8  \\
    \hline
    \end{tabular}%
  \label{first_crashes}%
\end{table}%
   	
\begin{table*}[h]
  \centering
  \caption{The Predicted Bug Locations of Deep Learning And The Real Bug Locations. Most real bug locations are successfully predicted by deep learning.}
 \resizebox{\textwidth}{!}{
    \begin{tabular}{|c|l|l|l|c|l|}
    \hline
    \multicolumn{4}{|c|}{Predicted bug locations by deep learning} & \multicolumn{2}{c|}{Real bug locations} \\
    \hline
    Library & \multicolumn{1}{c|}{Application} & File Name & Line Number & File Name & \multicolumn{1}{l|}{Line Number} \\
    \hline
    \multirow{7}{*}{libming} & listaction\_d & \multirow{7}{*}{decompile.c} & \multicolumn{1}{c|}{\multirow{7}{*}{103,370,381,407,440,455,476,477,555,569,583,714,726,762,1597,1690,1715,1932}} & \multicolumn{1}{l|}{decompile.c} & 440,455,370,381,714 \\
\cline{2-2}\cline{5-6}          & listswf\_d &       &       & \multicolumn{1}{l|}{decompile.c} & 370,440,407,477,455,381,476,1932,714,762,726,103 \\
\cline{2-2}\cline{5-6}          & swftocxx &       &       & \multicolumn{1}{l|}{decompile.c} & 1597,440,370,381,455,477 \\
\cline{2-2}\cline{5-6}          & swftoperl &       &       & \multicolumn{1}{l|}{decompile.c} & 407,440,370,455,381,477,714,1690 \\
\cline{2-2}\cline{5-6}          & swftophp &       &       & \multicolumn{1}{l|}{decompile.c} & 477,370,381,455,440,407,714,762,103,726 \\
\cline{2-2}\cline{5-6}          & swftopython &       &       & \multicolumn{1}{l|}{decompile.c} & 714,103,762,569,583,555,726,583 \\
\cline{2-2}\cline{5-6}          & swftotcl &       &       & \multicolumn{1}{l|}{decompile.c} & 381,440,477,455,370,1597,407,714,726,762,569,1715 \\
    \hline
    \multirow{2}{*}{gif} & \multirow{2}{*}{gifsponge} & egif\_lib.c & 92,764,802,1144 & egif\_lib.c & 771,790 \\
\cline{3-6}          &       & gifsponge.c & 44,76,81 & gifsponge.c & NA \\
    \hline
    \end{tabular}}%
  \label{bugs}%
\end{table*}%

The execution speed has a significant influence on fuzzing due to the random nature of fuzzing. With a higher execution speed, fuzzing saves time to run more executions. Fig.\ref{relative_speed} shows the relative execution speed of each application. For each fuzzer, we first calculate the average execution speed of each application. Then, the relative execution speed is the value that average execution speed of an individual fuzzer is divided by the average execution speed of both fuzzers.

As shown in Fig.\ref{relative_speed}, the relative execution speed of AFL is faster than DeFuzz on two applications, \texttt{Listaction\_d} and \texttt{Listswf\_d}. On the other hand, DeFuzz runs programs close or faster than AFL on ther other six applications. On the application \texttt{gifsponge}, although the execution speeds of the two fuzzers are close, AFL discovers only one path during the experiments and exposes zero crashes. On ther other hand, DeFuzz discovers much more crashes than AFL.
On average, the execution speed of AFL and DeFuzz is close to each other, and DeFuzz is slightly faster than AFL. Excluding \texttt{swftophp}, DeFuzz has a relative execution speed 4\% higher than AFL.

\begin{figure*}[!t]
		\centering
			\begin{minipage}[t]{0.9\textwidth}
				\centering
		        \includegraphics[width=\textwidth]{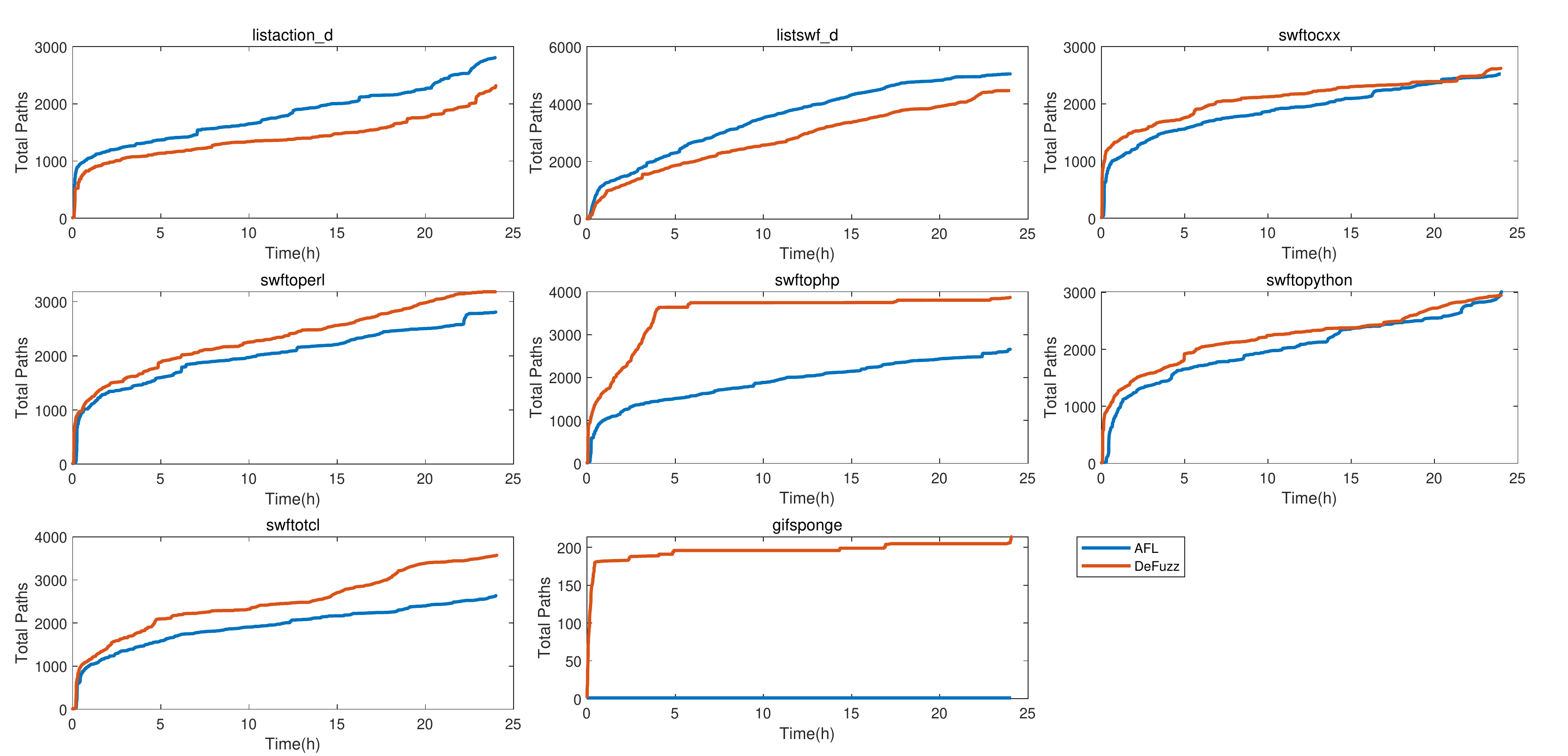}
			\end{minipage}%
		\caption{The number of paths. On average, DeFuzz discovers more paths than AFL.}
		\label{TotalPaths}%
	\end{figure*}
	
\begin{figure*}[h]
		\centering
			\begin{minipage}[t]{0.95\textwidth}
				\centering
		        \includegraphics[width=\textwidth]{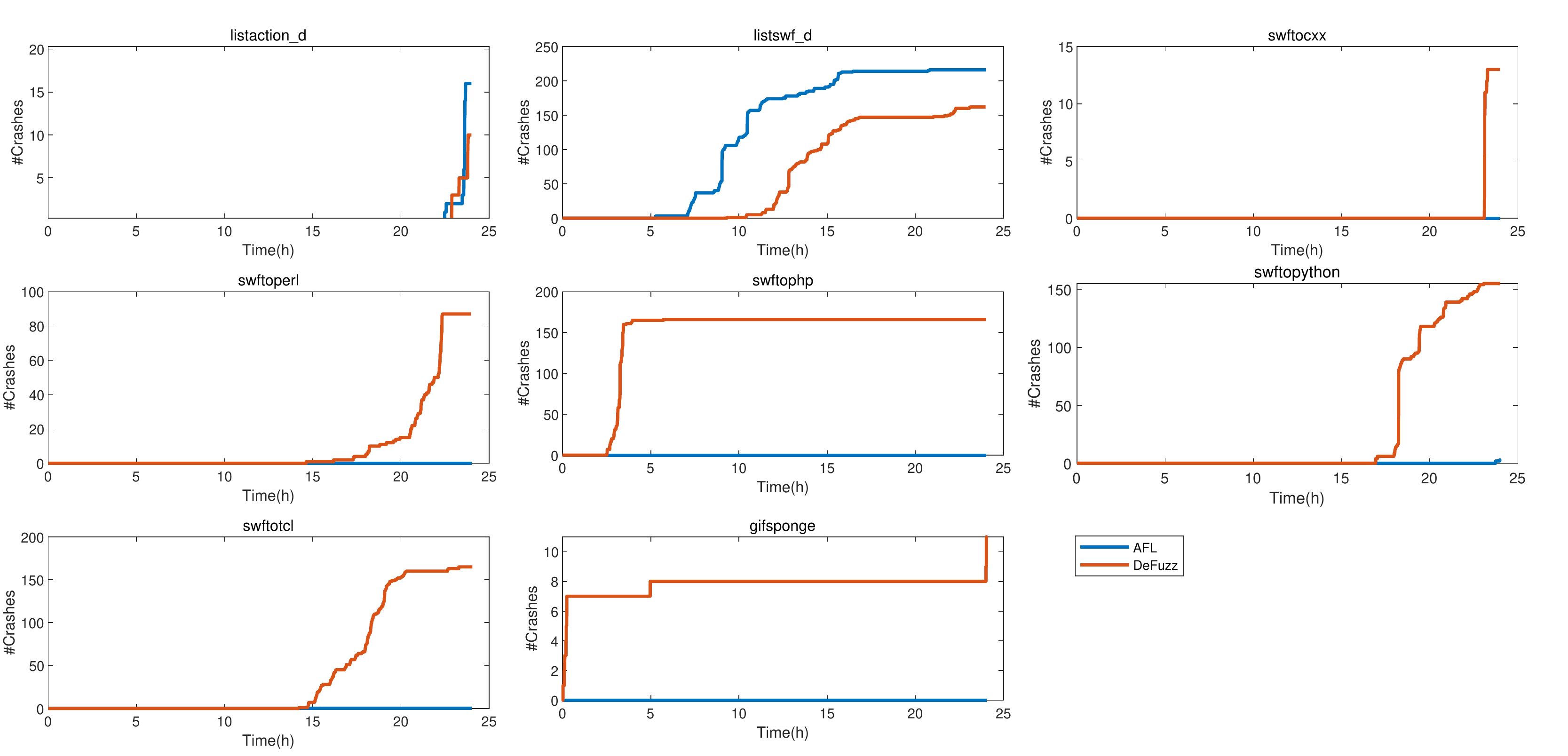}
			\end{minipage}%
		\caption{The number of crashes over time. DeFuzz exposes crashes faster than AFL.}
		\label{Crashes}%
	\end{figure*}

For coverage-guided fuzzing, such as AFL, it aims to discover as much coverage as possible. The assumption is that with a higher coverage, it has a higher chance to expose vulnerabilities. However, this might not always be true. Therefore, directed fuzzing, such as AFLGo, prefers to explore paths that are closer to the potential vulnerability locations. As DeFuzz is a directed fuzzer, it sets potential vulnerable locations utilising Deep Learning.

Fig.\ref{TotalPaths} shows that DeFuzz discovers more coverage than AFL on six applications except for \texttt{listaciton\_d} and \texttt{listswf\_d}. Although DeFuzz discovers less coverage than AFL on applications \texttt{listaciton\_d} and \texttt{listswf\_d}, DeFuzz discovers more crashes than AFL, as shown in Fig.\ref{Crashes}. Excluding  \texttt{gifsponge}, the overall code coverage of DeFuzz is only 4\% more than that of AFL, but DeFuzz detects three times as many crashes as AFL. Therefore, while DeFuzz is a directed fuzzing, it still discovers more coverage than AFL.

\subsection{Bug Discovery}

In this section, we will evaluate the performance of DeFuzz in terms of crash discovery and bug discovery.
The number of unique crashes is an important indicator of the effectiveness of fuzzing. The more crashes are detected, the greater probability of finding vulnerabilities.
TABLE \ref{crashes_num} shows that, within 24 hours, DeFuzz has successfully detected crashes on every target program. On the other hand, AFL did not detected any crash on the same applications except for \texttt{listaciton\_d} and \texttt{listswf\_d}. Therefore, DeFuzz is more effective than AFL to expose crashes.


We de-duplicated the crashes utilising \texttt{afl-collect}\cite{afl-collect}, and then verified bugs manually based on the crash stacks reproduced by GDB \cite{gdb}. 
The number of bugs discovered by fuzzers is shown in TABLE \ref{bugs-type}. Overall, DeFuzz discovers much more bugs than AFL on six applications. Meanwhile, the two fuzzers expose close number of bugs on applications \texttt{listaciton\_d} and \texttt{listswf\_d}. 
Note that we ran each trial for 24 hours, and AFL cannot expose crashes within 24 hours on six applications. We continued the experiments for AFL until it exposed crashes on the six applications that could not expose crashes within 24 hours. Table \ref{first_crashes} shows the time that AFL exposes the first crash on five applications. We could not find crashes on application \texttt{gifsponge}. 
The bug locations of each application are shown in TABLE \ref{bugs}, where the file name is the file in which bug happens and the line number is the corresponding line in the file.
The bug locations predicted by deep learning are also shown in TABLE \ref{bugs}. 

Comparing the real bug locations with the predicted ones, TABLE \ref{bugs} shows that deep learning has successfully predicted all the bug locations found by fuzzing on the seven applications from \texttt{libming}. On the application \texttt{gifsponge}, deep learning does not predict the accurate bug locations, however, the real bug locations are close to the predicted ones. For example, deep learning predicts that the bug locations include lines 764 and 802 in file \textit{egif\_lib.c} while the real bug locations are the lines 771 and 790. When DeFuzz can successfully guide fuzzing to the predicted locations, it has a high chance that the neighbour code lines can also be executed. Therefore, DeFuzz can discover bugs that are close to the predicted lines.

On the other hand, AFL only detects bugs on two applications \texttt{listaciton\_d} and \texttt{listswf\_d}. As shown in Table \ref{bugs-type}, DeFuzz found 63 bugs while AFL found only 20 bugs. Meanwhile, DeFuzz exposes two types of bugs, \textit{i.e.}, \texttt{out-of-memory} and \texttt{double free}. However, AFL only exposes bugs with the type \texttt{out-of-memory}.

\subsection{Limitations}

In order to guide fuzzing to target locations, DeFuzz has to calculate the distance between the current execution path and the target locations. However, the calculation of distance takes much time. Although DeFuzz calculates the distance during static analysis, which does not affect the execution speed of target programs, it may have problems when fuzzing on large programs. Moreover, DeFuzz uses control flow graph (CFG) to calculate distance. However, it is hard to statically construct an accurate CFG, which may miss some connections between two different basic blocks. The missed connections lead to an inaccurate distance.

Another limitation is that we usually face with the class imbalance problem with the data-driven cybersecurity. In this study, the ratio of vulnerable and not vulnerable samples we collected is about 1:16.5. In this case, the classifiers usually bias toward the majority class. For example, assume there are 95\% of the data samples are from majority class (e.g., not-vulnerable) while only 5\% of the data samples are from the minority class (vulnerable). If the classifier classify all the data samples as not-vulnerable, the accuracy will be 95\%, however, this is useless in the real-world security problem because we want to identify as many vulnerabilities as possible. Therefore, it is very necessary to address the class imbalance problem with data-driven cybersecurity problems. In our previous study, we developed DeepBalance \cite{liu2019deepbalance} to address the class imbalance problem, however, there are still much room for improvement. 

\section{Conclusion}
In this study, we proposed a DeFuzz, which is a deep learning guided directed fuzzing for software bugs identification. The goal is to use deep learning to identify the potential vulnerable locations, then use directed fuzzing to fuzz the vulnerable locations to reduce the false positive rate. To achieve this goal, DeFuzz uses the attention-based BiLSTM to train a prediction model based on the ground truth data we collected. The prediction model is used to identify any potential vulnerabilities given an unknown test case. We then run AFLgo to fuzz the potential vulnerable locations. We have demonstrated that DeFuzz outperforms the baselines by conducing experiments on real-world data.

Although DeFuzz achieves better performance that the baseline, there still several directions that one can further explore in the future. For example, deep learning can be utilised to guide fuzzing other than setting potential target locations. Deep learning can also be used to optimise the mutation scheme of fuzzing, such as how to mutate an input or the times an input to be mutated. 
Another interesting direction is to address the class imbalance problem when using the deep learning-based techniques.


\bibliographystyle{unsrt}
\bibliography{references}






\end{document}